%% file: main.tex
\pdfoutput=1
\PassOptionsToPackage{table}{xcolor}
\documentclass[sigconf]{acmart}
\usepackage{graphicx}
\usepackage{booktabs}
\usepackage{soul}
\usepackage{subfig}
\usepackage{xspace}
\usepackage{enumitem}
\usepackage{balance}
\usepackage{array}
\usepackage{lipsum}
\usepackage[export]{adjustbox}
\usepackage{caption}
\usepackage{mdframed}
\usepackage[table]{xcolor}

\newcommand{\hlc}[2][yellow]{{%
    \colorlet{foo}{#1}%
    \sethlcolor{foo}\hl{#2}}%
}

\AtBeginDocument{%
  \providecommand\BibTeX{{%
    \normalfont B\kern-0.5em{\scshape i\kern-0.25em b}\kern-0.8em\TeX}}}

\copyrightyear{2024}
\acmYear{2024}
\setcopyright{acmlicensed}
\acmConference[SIGIR '24]{Proceedings of the 47th International ACM SIGIR Conference on Research and Development in Information Retrieval}{July 14--18, 2024}{Washington, DC, USA}
\acmBooktitle{Proceedings of the 47th International ACM SIGIR Conference on Research and Development in Information Retrieval (SIGIR '24), July 14--18, 2024, Washington, DC, USA}
\acmDOI{10.1145/3626772.3657861}
\acmISBN{979-8-4007-0431-4/24/07}

\begin{document}
\sloppy

\author{Nandan Thakur}
\affiliation{University of Waterloo \city{Waterloo}\country{Canada}}\authornote{Corresponding author: <\url{nandan.thakur@uwaterloo.ca}>}

\author{Luiz Bonifacio}
\affiliation{UNICAMP and University of Waterloo \\ \city{Campinas}\country{Brazil}}

\author{Maik Fr\"obe}
\affiliation{Friedrich-Schiller-Universit\"at Jena \\ \city{Jena}\country{Germany}}

\author{Alexander Bondarenko}
\affiliation{Leipzig University and Friedrich-Schiller-Universit\"at Jena \\ \city{Leipzig}\country{Germany}}

\author{Ehsan Kamalloo}
\affiliation{University of Waterloo \city{Waterloo}\country{Canada}}

\author{Martin Potthast}
\affiliation{University of Kassel, hessian.AI, and ScaDS.AI \city{Kassel}\country{Germany}}

\author{Matthias Hagen}
\affiliation{Friedrich-Schiller-Universit\"at Jena \\ \city{Jena}\country{Germany}}

\author{Jimmy Lin}
\affiliation{University of Waterloo \city{Waterloo}\country{Canada}}

\settopmatter{authorsperrow=4}

\title{Systematic Evaluation of Neural Retrieval Models on the Touch{\'e}~2020 Argument Retrieval Subset of BEIR}


\renewcommand{\shortauthors}{Nandan Thakur et al.}

\begin{abstract}
The zero-shot effectiveness of neural retrieval models is often evaluated on the BEIR benchmark---a combination of different IR~evaluation datasets. Interestingly, previous studies found that particularly on the BEIR~subset Touch{\'e}~2020, an argument retrieval task, neural retrieval models are considerably less effective than~BM25. Still, so far, no further investigation has been conducted on what makes argument retrieval so ``special''. To more deeply analyze the respective potential limits of neural retrieval models, we run a reproducibility study on the Touch{\'e}~2020 data. In our study, we focus on two experiments: (i)~a black-box evaluation (i.e., no model retraining), incorporating a theoretical exploration using retrieval axioms, and (ii)~a data denoising evaluation involving post-hoc relevance judgments. Our black-box evaluation reveals an inherent bias of neural models towards retrieving short passages from the Touch{\'e}~2020 data, and we also find that quite a few of the neural models' results are unjudged in the Touch{\'e}~2020 data. As many of the short Touch{\'e} passages are not argumentative and thus non-relevant per se, and as the missing judgments complicate fair comparison, we denoise the Touch{\'e}~2020 data by excluding very short passages (less than 20 words) and by augmenting the unjudged data with post-hoc judgments following the Touch{\'e} guidelines. On the denoised data, the effectiveness of the neural models improves by up to~0.52 in~nDCG@10, but BM25 is still more effective. Our code and the augmented Touch{\'e}~2020 dataset are available at \url{https://github.com/castorini/touche-error-analysis}.
\end{abstract}

\ccsdesc[500]{Information systems~Retrieval models and ranking}
\ccsdesc[500]{Information systems~Evaluation of retrieval results}

\keywords{Argument retrieval; Neural retrieval models; Model evaluation}

\maketitle

\section{Introduction}
Substantial progress has been made in developing different types of neural retrieval models, including dense (e.g.,~\cite{yih:2011,lee:2019,karpukhin:2020,hofstatter:2021,xiong:2021}), sparse (e.g.,~\cite{dai:2021,zhao:2021,lin:2021,formal:2021}), and multi-vector models (e.g.,~\cite{khattab:2020,santhanam:2022,gao:2022,lee:2023}).
However, evaluations on the BEIR~retrieval benchmark~\cite{thakur:2021} show that the effectiveness of neural models substantially varies across different tasks and especially drops for some that lack dedicated training data (e.g., argument retrieval), while simple lexical BM25~retrieval tends to be more robust~\cite{thakur:2021}.
To address this problem, numerous efforts have spurred to improve the neural models' effectiveness by optimizing the training stage via knowledge transfer from high-resource datasets (e.g., MS~MARCO~\cite{nguyen:2016}), and with better mined hard negatives~\cite{formal:2021,ren:2021,santhanam:2022,ni:2022,asai:2023} by including an additional pretraining objective~\cite{izacard:2022,xiao:2022,gao:2022} or by using data augmentation via synthetic query generation~\cite{thakur:2021,thakur:2023,dai:2023}. 
Surprisingly, \emph{all} neural models continue to be less effective than BM25 on the Touch{\'e}~2020~\cite{bondarenko:2020} subset of BEIR, an argument retrieval task; cf.~\autoref{table-beir-effectiveness} with results for BM25 and state-of-the-art neural retrieval models like $\mathrm{E5_{large}}$~\cite{wang:2022b}, CITADEL+~\cite{li:2023}, SPLADEv2~\cite{formal:2021}, etc.

\input{tables/table-beir-effectiveness}

Motivated by this observation, we conduct a two-stage reproducibility study on the Touch{\'e}~2020 data to understand the potential respective limits of current neural retrieval models.
Our first stage are black-box evaluations (i.e., without requiring model retraining) to examine and possibly somewhat correcting errors incurred by the neural models. In first analyses, we find that the neural models on average retrieve much shorter arguments than~BM25  (Sections~\ref{sec:black-box-model-evaluation}). For instance, about half of the top-10 results of dense retrievers (e.g., TAS-B~\cite{hofstatter:2021}) contain at most two sentences that often are not even argumentative (e.g., ``Pass'' or ``I agree with lannan13'') yielding low effectiveness scores.
To possibly improve the neural model's effectiveness, we then repeat the evaluation on augmented versions of the Touch{\'e}~2020 corpus: (i)~via document expansion%
\footnote{In argument retrieval, the terms `argument' and `document' are used interchangeably.}
using DocT5query~\cite{nogueira2019doc2query} (lengthening short arguments) and (ii)~via document summarization with GPT-3.5~\cite{ouyang:2022} (shortening longer arguments). The corpus augmentation does not require a retraining of the neural models and our results show that the effectiveness indeed increases for a majority of the models (Section~\ref{sec:inference-time-features}).

In our second reproducibility stage, we analyze intrinsic characteristics of the Touch{\'e}~2020 corpus and find that, unlike for other BEIR~subsets, about 20\% of the Touch{\'e}~2020 corpus are very short documents (and thus mostly non-argumentative) and that at least 50\%~of the documents retrieved by neural models (and even BM25) are actually unjudged and thus considered non-relevant in standard evaluation setups. 
To counter these effects, we carefully remove short documents (less than 20~words) from the Touch{\'e}~2020 corpus (Section~\ref{sec:denoising}) and we add missing judgments following the Touch{\'e} guidelines (Section~\ref{sec:post_hoc_judgements}). Our experimental results show that without very short documents in the corpus and with added post-hoc judgments, the effectiveness of all neural models substantially improves by up to~0.52 in terms of nDCG$@$10. Yet, even after denoising and post-hoc judgments, BM25 remains the most effective.

We finally supplement our findings with a theoretical analysis using information retrieval axioms~\cite{bondarenko:2022} on the Touch{\'e}~2020 data (Section~\ref{sec:axioms}) and find that all neural models violate the document length normalization axiom LNC2~\cite{fang:2011}, which is well-supported by BM25. Overall, our contributions are the following:

\begin{itemize}[leftmargin=*]
\item We reproduce dense, sparse, and multi-vector neural retrieval models on the BEIR~subset Touch{\'e}~2020 (argument retrieval) and find that short and low-quality arguments substantially harm the effectiveness of many neural models.
\item After carefully denoising the Touch{\'e}~2020 corpus and adding post-hoc judgments, the effectiveness of all neural models substantially improves. However, BM25 remains more effective.
\item Our code and the denoised, post-hoc judged dataset are available at \href{https://github.com/castorini/touche-error-analysis}{https://github.com/castorini/touche-error-analysis}.
\end{itemize}

\section{Background and Related Work}

Argument retrieval is the task of ranking documents based on the topical relevance to argumentative queries (i.e., queries about debated topics like ``Should bottled water be banned?''), i.e., the documents should contain appropriate arguments pertinent to the query. 
An argument is often modeled as a conclusion (i.e., a claim that can be accepted or rejected) and a set of supporting or attacking premises (i.e., reasons to accept or reject the conclusion like statistical evidence, an anecdotal example, etc.)~\cite{wachsmuth:2017b, stab:2018a}.

Previous works on argument retrieval~\cite{potthast:2019,shahshahani:2020} majorly made use of lexical retrieval models such as BM25~\cite{robertson:1994}, DirichletLM~\cite{zhai:2017}, DPH~\cite{amati:2006}, and TF-IDF~\cite{jones:2004}. These models were also commonly used to retrieve argumentative documents in argument search engines. 
For instance, popular argument search engines such as args.me~\cite{wachsmuth:2017b}, ArgumenText~\cite{stab:2018a}, and TARGER~\cite{chernodub:2019}, all utilize BM25 for retrieving argumentative documents. 
Further, a large body of work to study argument retrieval approaches was carried out as part of the Touch{\'e}'s shared task on argument retrieval for controversial questions~\cite{bondarenko:2020}. 
Most of the submitted approaches by the task participants also used lexical retrieval models (e.g., BM25 and DirichletLM) for document retrieval combined with various query processing, query reformulation, and expansion techniques. In our work, we focus on evaluating neural retrieval models as lexical retrieval models have already been well examined and utilized in argument retrieval.

The Touch{\'e}~2020 dataset (queries, document collection, and relevance judgments) was later included as an argument retrieval subset in the BEIR benchmark for zero-shot evaluation of neural retrieval models in \citet{thakur:2021}. 
Interestingly, none of the tested neural retrieval models, trained on MS~MARCO~\cite{nguyen:2016}, outperform BM25 on the Touch{\'e}~2020 argument retrieval task, as shown in Table~\ref{table-beir-effectiveness}. 
But neural models outperform BM25 on a majority of the other datasets included in the BEIR benchmark (e.g., MS~MARCO~\cite{nguyen:2016} or Natural Questions~\cite{kwiatkowski:2019}). 
Subsequent works improving model generalization on BEIR such as $\mathrm{E5_{large}}$ \cite{wang:2022b}, CITADEL+ \cite{li:2023} or DRAGON+ \cite{lin:2023} continue to underperform on Touch{\'e}~2020.

The study in \citet{thakur:2021} was one of the earliest works to observe the tendency of dense retrievers to retrieve short documents in Touch{\'e}~2020 and provided a theoretical explanation using different similarity measures in the training loss function. 
In our work, we extend the idea from \citet{thakur:2021} and conduct a more thorough systematic evaluation by including diverse neural model architectures and examining the Touch{\'e}~2020 corpus.

Prior works have suggested several ways to understand the relationship between retrieval effectiveness and quality of test collections via empirical analyses. 
For instance, train--test leakage~\cite{lewis:2021}, retrievability bias due to query length~\cite{wilkie:2015}, sampling bias due to near-duplicates~\cite{frobe:2020}, or saturated leaderboards unable to distinguish any meaningful improvements~\cite{arabzadeh:2022} were examined. 
However, prior work has missed out on evaluating the impact of document corpora on retrieval effectiveness, i.e., the potential impact of non-relevant documents present within a corpus on neural models. In our work, we conduct a comprehensive evaluation by independently evaluating both the Touch{\'e}~2020 dataset and retrieval models to help devise targeted strategies for model improvement or data cleaning.

\section{Experimental Setup}\label{sec:experimental-setup}

In this section, we review the Touch{\'e}~2020 dataset used for argument retrieval and provide details on the baseline retrieval models. Next, we provide details on model evaluation and implementation.

\paragraph{Touch{\'e}~2020} The Touch{\'e}~2020 task on controversial argument retrieval~\cite{bondarenko:2020} uses a focused crawl of arguments for 49~test queries addressing socially important (and often controversial) issues like ``Should bottled water be banned?''. The document collection is the args.me corpus~\cite{ajjour:2019} containing 382,545~arguments. Each argument has a title in the form of a conclusion (i.e., a claim that an arguer could make) and a context containing several premises (reasons, opinions, or evidence that support or attack the claim). The task organizers also published relevance judgments (non-relevant, relevant, and highly relevant) for 2,214~documents (cf.\ Table~\ref{tab:touche-stats} for dataset characteristics). The documents were pooled using the top-5 pooling strategy from 12~ranked results submitted by participants. While the argument retrieval track in Touch{\'e} has more recent versions \cite{bondarenko:2021, bondarenko2022b, bondarenko:2023}, in our work, we use the Touch{\'e}~2020 dataset, due to its availability in the BEIR benchmark \cite{thakur:2021}.%
\footnote{\href{https://public.ukp.informatik.tu-darmstadt.de/thakur/BEIR/datasets/webis-touche2020.zip}{ukp.informatik.tu-darmstadt.de/thakur/BEIR/datasets/webis-touche2020.zip}}

\paragraph{Retrieval Models} For our experiments, we select different open-source retrieval models to better understand errors across different neural architectures. The selected models are the lexical model BM25~\cite{robertson:1994}, the dense retrievers DRAGON+~\cite{lin:2023}, Contriever~\cite{izacard:2022}, and TAS-B~\cite{hofstatter:2021}, the sparse retriever SPLADEv2~\cite{formal:2021}, and the multi-vector retriever CITADEL+~\cite{li:2023}.

\begin{figure}[tb]
    \centering
    \includegraphics[width=1.\columnwidth, trim=12 10 25 0,clip]{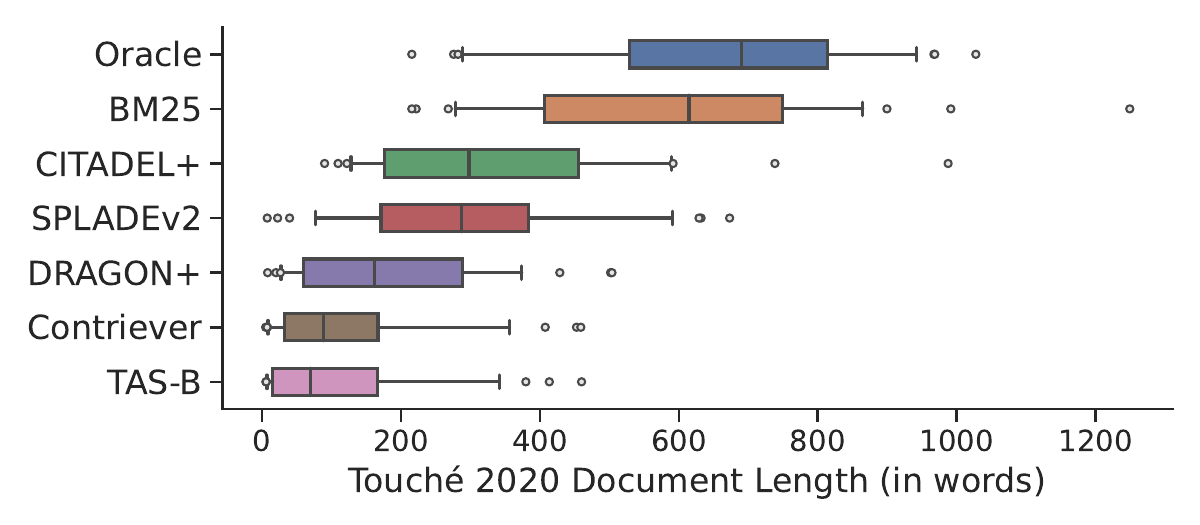}
    \caption{Boxplots showing the average length in words ($x$-axis) of the top-10 Touch{\'e}~2020 results retrieved by the models on the $y$-axis (sorted by decreasing nDCG@10; oracle: avg.\ length of all documents judged as relevant). The results of the neural models are much shorter in comparison to BM25.}
    \label{fig:boxplot-avg-doc-length}
\end{figure}

\paragraph{Evaluation} To evaluate retrieval effectiveness on Touch{\'e}~2020, we use nDCG@10 metric as it has been widely adopted in the BEIR benchmark~\cite{thakur:2021}. In addition, we use the hole$@k$ rate (i.e., the ratio of results retrieved by a model at cutoff~$k$ that do not have relevance judgments) to estimate the proportion of unjudged documents. 

\paragraph{Implementation Details}
In our work, we conduct a reproducibility study with previously available models' checkpoints. We did not retrain any neural model and use up to a maximum of A6000~$\times$~4 GPUs for inference.
For BM25, we follow \citet{thakur:2021} and use multi-field (title and body indexed separately with equal weights) version%
\footnote{\href{https://github.com/castorini/anserini}{https://github.com/castorini/anserini}}
available in Anserini~\cite{yang:2016} with default parameters ($k_1=0.9$ and $b=0.4$). 
For our dense models, Contriever (mean pooling with dot product), TAS-B, and DRAGON+ (both \texttt{[CLS]} token pooling with dot product), we reproduce the results by converting model checkpoints using \texttt{sentence-transformers}\footnote{\href{https://github.com/UKPLab/sentence-transformers}{https://github.com/UKPLab/sentence-transformers}} and evaluate them on Touch{\'e}~2020 using BEIR evaluation.%
\footnote{\href{https://github.com/beir-cellar/beir}{https://github.com/beir-cellar/beir}}
For SPLADEv2 (max aggregation), we reproduce the model using the SPRINT toolkit~\cite{thakur:2023}.%
\footnote{\href{https://github.com/thakur-nandan/sprint}{https://github.com/thakur-nandan/sprint}}
Finally, for CITADEL+ (with distillation and hard negative mining), we use the original \texttt{dpr-scale} repository for reproduction.%
\footnote{\href{https://github.com/facebookresearch/dpr-scale/tree/citadel}{https://github.com/facebookresearch/dpr-scale/tree/citadel}}
Apart from DRAGON+, in our work, we successfully reproduce the nDCG@10 on Touch{\'e}~2020.\footnote{For DRAGON+, we suspect the difference being caused by using A100 vs.\ A6000~GPUs.}

\section{Evaluation Experiments}

In this section, we describe our evaluation experiments consisting of two independent parts. 
First, we conduct a black-box evaluation to understand the limitations of neural models on Touch{\'e}~2020 (Section~\ref{sec:black-box-model-evaluation}) and propose two methods to improve the neural model effectiveness at inference time (Section~\ref{sec:inference-time-features}). 
Next, we denoise the data by filtering out short documents (Section~\ref{sec:denoising}) and conduct post-hoc relevance judgments (Section~\ref{sec:post_hoc_judgements}) to measure the unbiased nDCG@10 of neural models versus BM25 on Touch{\'e}~2020. 
Finally, we attempt to theoretically understand our findings using axioms for information retrieval (Section~\ref{sec:axioms}).

\input{tables/retrieval-document-examples}

\subsection{Black-Box Model Evaluation on Touch{\'e}~2020}\label{sec:black-box-model-evaluation}

The neural retrieval model's training often involves one or several of the following steps, a particular training dataset selection~\cite{nguyen:2016,kwiatkowski:2019}, choosing a training optimization objective~\cite{karpukhin:2020,hofstatter:2020} and deciding whether to train with specialized hard negatives~\cite{qu:2021,hofstatter:2021}. These configurations are crucial for neural model effectiveness but lack explainability. Hence, our objective is to uncover the reasons for errors of retrieval models (BM25 vs.\ neural models) on Touch{\'e}~2020, by treating models as black-boxes (without modifying parameters). Specifically, we ask the following research question:

\newcommand{\RQone}{\begin{itemize}
    \item[\textbf{RQ1}] \emph{Does the non-uniformity in document lengths affect neural model effectiveness on the Touch{\'e}~2020 dataset?}
\end{itemize}}
\RQone

\paragraph{Quantitative Results} 
Figure~\ref{fig:boxplot-avg-doc-length} shows boxplots depicting the average document lengths of the top-10 retrieved documents by the models under investigation, where the whiskers plot the 95\% confidence interval. 
The lengths are computed as word counts after applying \texttt{nltk} word tokenizer~\cite{bird:2009}.
All neural models, on average, retrieve shorter documents containing less than 350 words (visible from medians and whiskers in Figure~\ref{fig:boxplot-avg-doc-length}) in contrast to BM25, which retrieves longer documents on average containing more than 600 words which best mimics the Oracle distribution. Dense models (TAS-B, Contriever, and DRAGON+) appear to retrieve the shortest arguments, followed by sparse (SPLADEv2) and muti-vector (CITADEL+).
The decrease in nDCG@10 on Touch{\'e}~2020 is found to be \emph{perfectly correlated} with the increase in shorter top-10 retrieved documents (Spearman correlation $\rho=1.0$). Overall, this provides positive evidence for our hypothesis that the shorter documents present in Touch{\'e}~2020 (cf.\ Figure~\ref{fig:document-length-distribution}) negatively affect neural models in terms of retrieval effectiveness.

\input{tables/error-rate-table}

\paragraph{Empirical Evidence} 
Upon a careful analysis of the retrieved documents by the models under investigation, we observe an interesting pattern across the retrieved documents.
We find that documents retrieved in Touch{\'e}~2020 by neural models show a high overlap of the query terms with the argument conclusion (document title) which is often relevant, but includes a rather ``noisy'', i.e., short argument premise (document body) which is non-relevant, e.g., a single word ``Pass'' or ``social security is in crisis'' (an example for a test query is shown in Table~\ref{tab:retrieval-documents-examples}). 
To quantify the empirical evidence, we compute an error rate (in~\%) by counting a mistake the model makes if the document retrieved (i)~is non-relevant (relevance either 0 or unjudged), and (ii)~is shorter than 1--2 sentences (a maximum of 20 words).
From Table~\ref{tab:retrieval-documents-examples}, we observe that dense retrievers suffer the most with TAS-B with the highest 51.6\% error rate in top-10 retrieved documents. CITADEL+ contains a lower percentage of shorter non-relevant documents with a low~4.5\% error rate. BM25 has the lowest error rate of~0.8\%, which suggests that BM25 is empirically found to be robust against non-uniformity in document lengths present within the Touch{\'e}~2020 corpus.

\begin{figure}[t!]
\begin{mdframed}[backgroundcolor=gray!5]
    \small
    \texttt{system: \\
You are an Argument Summarizer, an intelligent assistant that can summarize an argument. The output summary must be written using an argument nature.\\\\
    assistant: \\
    Okay, please provide the argument. \\\\
    user:\\
    \{argument\_text\}
}
\end{mdframed}
\caption{Vanilla zero-shot prompt template used in our work with GPT-3.5 \cite{ouyang:2022} to summarize Touch{\'e}~2020 documents.}
\label{fig:prompt_gpt4}
\end{figure}

\begin{figure}[t]
    \centering
    \includegraphics[trim=0 10 0 0,clip,width=0.45\textwidth]{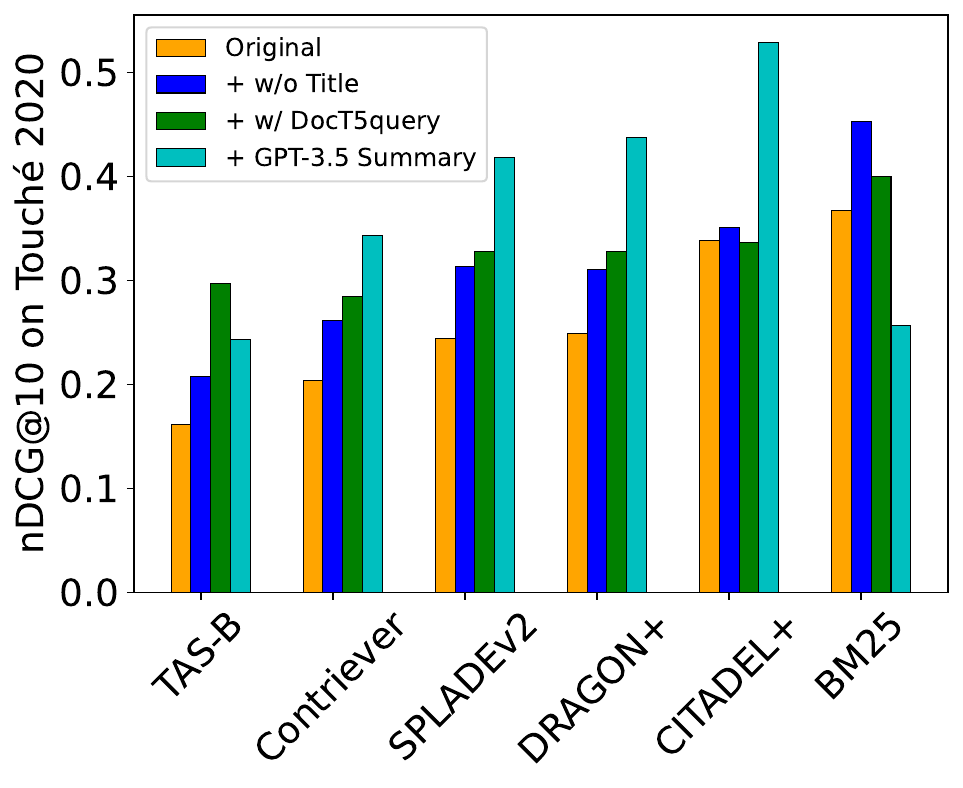}%
    \caption{%
       Change in effectiveness with DocT5query \cite{nogueira2019doc2query} query expansions and GPT-3.5 \cite{ouyang:2022} summary replacement on Touch{\'e}~2020. Both techniques improve the nDCG@10 for a majority of the neural models. 
       }%
    \label{fig:results-exp-summ}
\end{figure}

\paragraph{Reasoning} 
We hypothesize reasons for the observed error pattern. We start by assuming query $q_i$ and the short non-relevant document $\hat{d}_i$ have a high word overlap due to similarity with the conclusion (title), i.e., the document is a good paraphrase of the query but does not contain information to answer the question. 
As shown previously in \citet{ram:2023}, lexical overlap remains a highly dominant signal for relevance in dense retrievers, which we suspect causes the non-relevant short document, with a similar length to the query, closer within the dense embedding space representation. 
For sparse and multi-vector retrievers, the token overlap of query $q_i$ and the shorter non-relevant document is high, which results in a higher similarity score. 
However, in contrast, the BM25 algorithm accounts for document length normalization within its parameter~$b$~\cite{singhal:1996}.
As longer documents tend to have more term occurrences, leading to potential bias, document length normalization in BM25 acts as a normalization parameter, improving robustness against sensitivity toward short document errors.
Neural models, conversely, suffer from noise present in the form of short and non-relevant documents in Touch{\'e}~2020.

\subsection{Improving Effectiveness at Inference Time}\label{sec:inference-time-features}

Information retrieval datasets such as Touch{\'e}~2020 (unlike MS MARCO or Natural Questions), may not be uniform in document length. 
Ideally, models should be explicitly trained to be robust against noisy short documents, but practitioners lack access to these setups, and retraining is often computationally expensive. Based on these observations, we ask the following research question:

\newcommand{\RQtwo}{\begin{itemize}
    \item[\textbf{RQ2}] \emph{Can we improve neural model effectiveness at inference time without expensive retraining of models?}
\end{itemize}}
\RQtwo

We experiment with two techniques to improve neural model effectiveness at inference time: (i)~expanding documents with synthetic DocT5query queries, and (ii)~shortening documents by replacing them with GPT-3.5-generated summaries.

\input{tables/exp-summ-final}

\paragraph{DocT5query Expansion} 
We reuse the DocT5query \cite{nogueira2019doc2query} model\footnote{\href{https://huggingface.co/BeIR/query-gen-msmarco-t5-base-v1}{https://huggingface.co/BeIR/query-gen-msmarco-t5-base-v1}} from BEIR \cite{thakur:2021} to expand documents in Touch{\'e}~2020 with generated queries. 
We focus solely on generating queries using the premise (body) and not the conclusion (title) for all 382,545~documents in Touch{\'e}~2020. 
We hypothesize that noisy, shorter documents that negatively affect retrievers will increase the document length and decrease relevance as they now contain additional non-relevant terms (generated queries would repeat these terms).
For our experiments, we generate 10~synthetic queries for each document within Touch{\'e}~2020, append these synthetic queries to their respective argument (cf.\ Table~\ref{tab:examples-summ-exp}), and re-evaluate all tested models.

\paragraph{GPT-3.5 Summarization} 
Furthermore, we explore an additional technique by using shorter and relevant summaries to replace lengthy documents in Touch{\'e}~2020.
Using GPT-3.5-turbo \cite{ouyang:2022}, we generate concise summaries of all the 2,214 originally judged documents in Touch{\'e}~2020 available as a proxy\footnote{Generating summaries using GPT-3.5 for all 382,545 documents in Touch{\'e}~2020 is expensive and not feasible within our computational budget.} using a zero-shot vanilla prompt template shown in Figure~\ref{fig:prompt_gpt4}. We replace the original judged document with the summarized version. The synthetic summaries typically follow a uniform structure, starting with an introductory overview of the topic, followed by supporting or opposing premises, with examples and evidence originally discussed in the source document (cf.\ Table~\ref{tab:examples-summ-exp}).

\paragraph{Experimental Results} 
As shown in Figure~\ref{fig:results-exp-summ}, removing conclusions (document titles) from arguments improves the nDCG@10 on Touch{\'e}~2020 across all models, with a particularly pronounced effect on BM25. We discuss more about this later in Section~\ref{sec:denoising}. The DocT5query-based expansion improves TAS-B on Touch{\'e}~2020 with minor improvements for other neural models, except CITADEL+. As hypothesized, document expansion with generated queries helps neural models to smartly avoid retrieving short and non-relevant documents by extending them with additional non-relevant terms (see Table~\ref{tab:examples-summ-exp} for reference). With GPT-3.5 replaced summaries, BM25 shows a decline in nDCG@10, whereas other neural models like DRAGON+ and CITADEL+ show significant improvements in nDCG@10 on Touch{\'e}~2020. 
The absence of query terms in the GPT-3.5 summary may impact BM25's ability to effectively match query terms, unlike neural models' semantic representation, which can fit more relevant information within their (maximum) sequence length constraint of 512 tokens, thereby helping neural models to retrieve better documents as summaries.

\subsection{Denoising the Touch{\'e}~2020 Corpus}\label{sec:denoising} 
\begin{figure}[t!]
\centering
\subfloat[MS MARCO~\cite{nguyen:2016}]{
\includegraphics[width=0.23\textwidth, trim={0 0 0 0}]{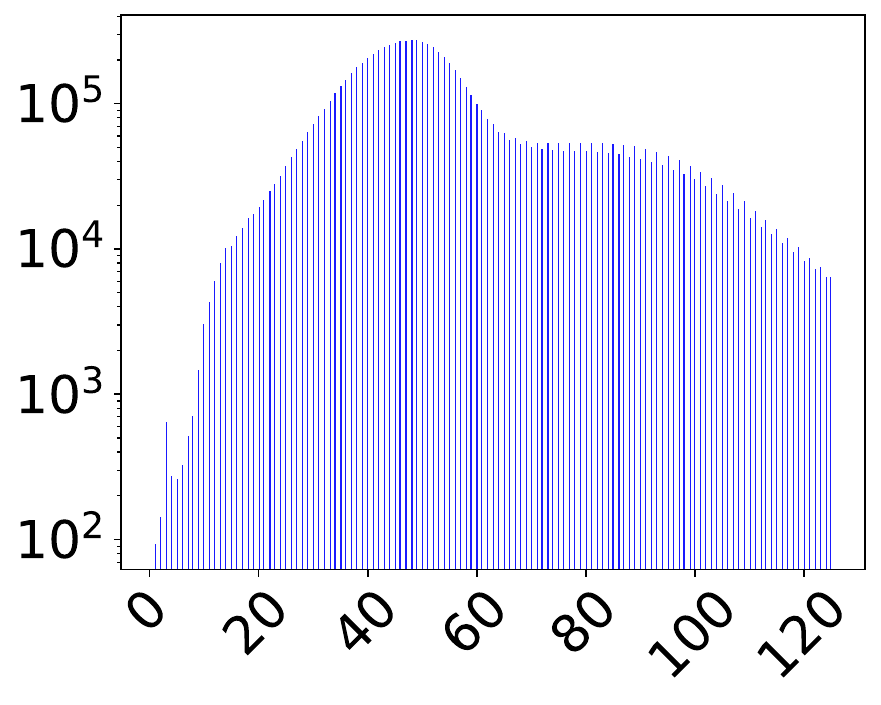}
\label{fig:msmarco-distribution}
}
\subfloat[Touch{\'e}~2020~\cite{bondarenko:2020}]{
\includegraphics[width=0.23\textwidth, trim={0 0 0 0}]{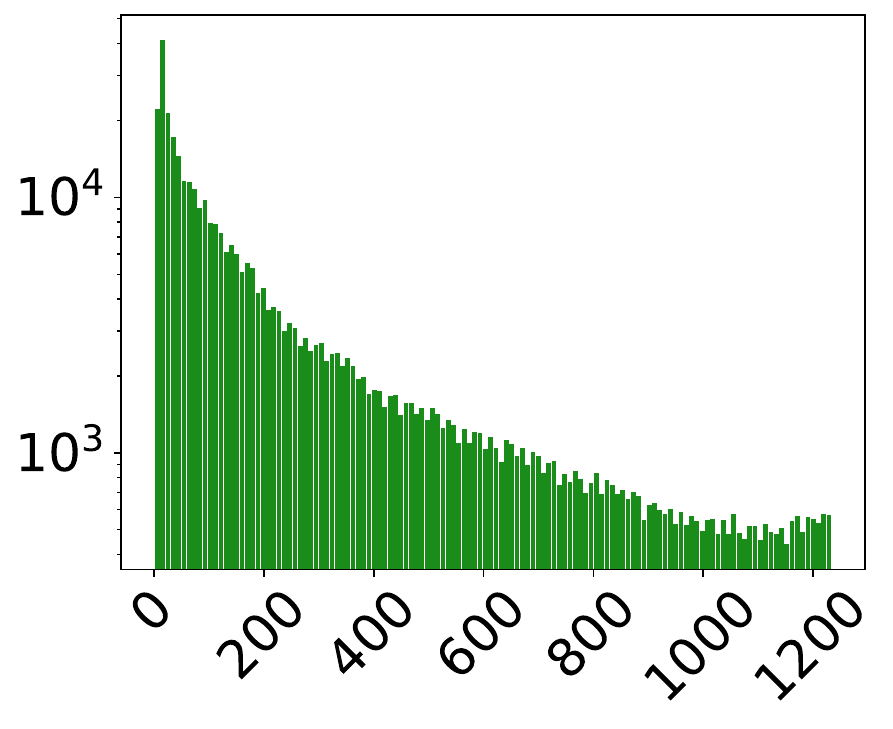}
\label{fig:touche-distribution}
}
\caption{Document length distribution in Touch{\'e}~2020 vs.\ MS MARCO ($x$-axis: document length in words; log-scaled $y$-axis: frequency of document lengths). Touch{\'e}~2020 has a monotonically decreasing broad distribution, while the MS~MARCO distribution is much narrower.}
\label{fig:document-length-distribution}
\end{figure}

As discussed in Section~\ref{sec:experimental-setup}, the args.me corpus in Touch{\'e}~2020 contains web-crawled arguments from various debating portals and thereby may contain noise as non-valid arguments. 
However, a valid document premise (or body) should provide evidence or reasoning that can be used to back up the conclusion (or title) as an argument \cite{stab:2018b,toledo:2019,sun:2021}. But very short premises that are less than 1--2 sentences (e.g., ``Pass'' or ``I agree'') do not contain enough evidence to be classified as a valid argument.

To better understand the Touch{\'e}~2020 corpus, we compare its document length distribution to the standard retrieval dataset MS~MARCO~\cite{nguyen:2016}.
The plots in Figure \ref{fig:document-length-distribution} show that the length distribution of Touch{\'e}~2020 monotonically decreases with a high frequency of extremely short arguments (spike in the graph at 20--30~words) and a long tail of long arguments (even exceeding 1200~words), while the MS~MARCO length distribution is much narrower with relatively few extremely short outliers.

\begin{figure}[t]
    \centering
    \includegraphics[trim=0 10 0 0,clip,width=0.4\textwidth]{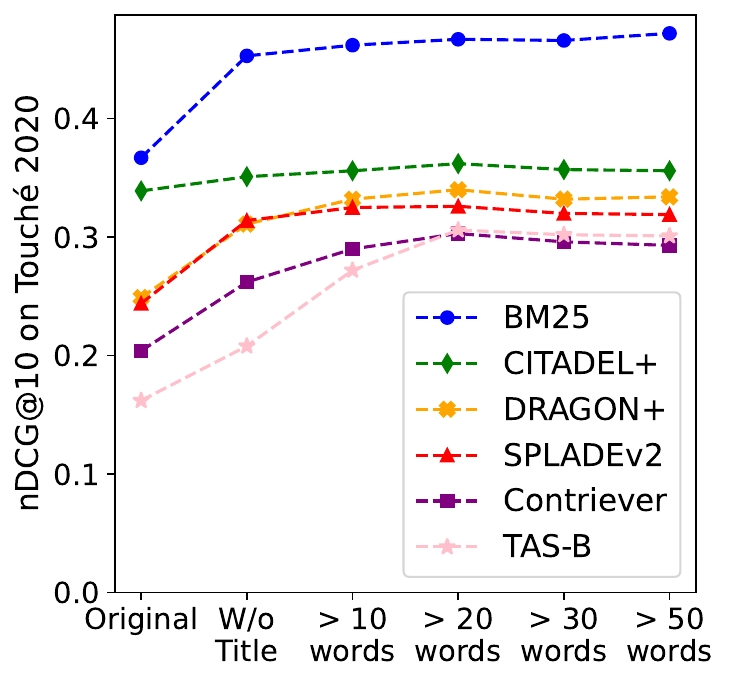}%
    \caption{%
    Denoising experiment to determine the best threshold $n$ for filtering out short documents in Touch{\'e}~2020. All models improve (until a maximum of 20 words) in effectiveness with data denoising in Touch{\'e}~2020.}\label{fig:dataset-filteration} 
\end{figure}

\newcommand{\RQthree}{\begin{itemize}
    \item[\textbf{RQ3}] \emph{Does neural retrieval model effectiveness improve by denoising the Touch{\'e}~2020 document corpus?}
\end{itemize}}
\RQthree

The hypothesis for investigating this research question lies in whether the effectiveness of neural models can be improved by cleaning, i.e., reducing noise in the Touch{\'e}~2020 document corpus. To validate this, we experiment by reducing noise in Touch{\'e}~2020, i.e., filtering out non-argumentative documents from the corpus. One way is to use argument classification to classify each document~\cite{reimers:2019,daxenberger:2020} as either a valid or non-valid argument, however, it is computationally expensive to classify all arguments in Touch{\'e}~2020~\cite{reimers:2019,gienapp:2020}. Instead, we follow a simple heuristic and filter out potentially non-valid arguments based on the document length. Our denoising technique removes the conclusion (across all documents in Touch{\'e}~2020) and only carefully selects documents with premises greater than a threshold of at least $n$~words in length.

\paragraph{Results after Denoising} 
Figure~\ref{fig:dataset-filteration} shows that our heuristic denoising improves the nDCG@10 for all models. That removing the argument conclusion (i.e., title) alone improves the nDCG@10 for all models is probably caused by the inherent nature of argument retrieval, where premises are more important for a document to be classified as a valid argument than the conclusion. Without the conclusion, often also the lexical overlap with the query that confuses neural models (cf.~Section~\ref{sec:black-box-model-evaluation}) is decreased.
As for a length threshold for removing documents, $n=20$~words empirically provides the best nDCG@10 across all tested models, as the effectiveness saturates when removing premises with more than 20~words.

A limitation of denoising Touch{\'e}~2020 is that we miss out on a few human-judged query-document pairs with document lengths shorter than 20 words. However, as Table~\ref{tab:touche-stats} shows, overall~89\% (382 out of 429) of the missed judgments were originally non-relevant (score~0), and only~3.7\% (16 out of 429) are highly relevant (score~2). This suggests that shorter documents in the Touch{\'e}~2020 corpus are likely to be non-relevant, hence denoising based on document length is a good and simple heuristic for checking valid arguments in the argument retrieval task.

\subsection{Adding Post-hoc Relevance Judgments}\label{sec:post_hoc_judgements}
Retrieval datasets can contain multiple biases induced by either the annotation guidelines, annotation setup, or human annotators. For instance, to avoid selection bias~\cite{lipani:2018} in later studies using some retrieval dataset, popular information retrieval challenges, for instance at TREC~\cite{craswell:2024, lawrie:2022}, aim to encouragethe submission of diverse retrieval approaches to yield diverse judgment pools.

\input{tables/dataset-filteration}

To quantify the selection bias in Touch{\'e}~2020, we compute how many of the top-10 results of our tested models are unjudged in the original and denoised corpus versions. Table~\ref{tab:filtered-results} shows that the respective hole$@$10 values all are greater than~50\% (i.e., more than half of the top results of every model are unjudged in the Touch{\'e}~2020 data). Therefore, we ask the following research question: 

\newcommand{\RQfour}{\begin{itemize}
    \item[\textbf{RQ4}] \emph{Are neural retrieval models unfairly penalized  on Touch{\'e}~2020 due to a selection bias?}
\end{itemize}}
\RQfour

\paragraph{Annotation Details} We conduct a post-hoc relevance judgment study to fill up the hole@10 across all tested models, i.e., annotating originally unjudged arguments, as filling up holes would account for denser judgments and a better estimate of nDCG@10. We hired 5$+$ annotators with prior debating experience and follow annotation guidelines available in \citet{bondarenko:2020}. We conduct the post-hoc judgments and fill up hole@10 for all tested models by evaluating each unjudged document with three relevance labels: 0~(non-relevant), 1~(relevant), and 2~(highly relevant). We cumulatively took around 10--15 hours to judge 1,064 judgment pairs and paid each annotator a competitive hourly rate of 14.86 USD per hour.
Table~\ref{tab:touche-stats} contains Touch{\'e}~2020 statistics before and after the denoising and post-hoc judgment rounds. In our post-hoc judgment round, over 78\%~of the judgment pairs were judged relevant (with 48\%~highly relevant and 30\%~relevant), indicating that many ``relevant'' documents are retrieved by models but unjudged originally in Touch{\'e}~2020.
We measure the inter-annotator agreement score with Fleiss'~$\kappa$~\cite{fleiss:1971}. Since argument retrieval is highly subjective and biased towards annotator preferences and beliefs as discussed in \cite{bondarenko:2020,hua:2019}, we achieve a comparable score of $\kappa=0.31$.\footnote{We earlier observed a lower $\kappa$ due to mistakes from a single annotator, which we discussed internally and rectified.}

\input{tables/filteration-results}

\paragraph{Results after Post-hoc Judgments} 
Re-evaluation scores of the retrieval models after post-hoc judgment rounds are shown in Table~\ref{tab:filtered-results} (column `++~Post-hoc').
The maximum increase in nDCG$@$10 is observed in dense retrievers (TAS-B, Contriever, and DRAGON+) and the least in multi-vector retrieval with CITADEL+. 
This provides evidence that post-hoc relevance judgments to fill up holes are necessary for a fair evaluation of models. 
In our hypothesis, we suspected a bias towards lexical retrievers due to their dominance in the original candidates during original Touch{\'e}~2020 judgment rounds. 
However, even after post-hoc relevance judgments with more and better ``semantic'', i.e., neural retrieval models, and denoising Touch{\'e}~2020, BM25 continues to outperform all neural models by a margin of at least 6.7 points on nDCG@10, thereby making it still a robust baseline for argument retrieval. 

\subsection{Axiomatic Error Analysis on Touch{\'e}~2020} \label{sec:axioms}

To contrast our previous empirical evaluation of neural retrieval models on Touch{\'e}~2020 with well-grounded theoretical foundations of information retrieval, we investigate if we can observe similar trends using axiomatic analysis. Therefore, we measure the agreement of the neural models under investigation with information retrieval axioms.
A higher agreement indicates that a retrieval model fulfills the theoretical constraint introduced in the axiom. 
These axioms can highlight the problems in neural models, and fixing these problems can improve the model's effectiveness~\cite{bondarenko:2022}, even when there is no strong correlation between axioms and relevance judgments~\cite{camara:2020}.
While retrieval axioms can increase the effectiveness of neural retrieval models (e.g., when used for regularization~\cite{rosset:2019}), dedicated axioms for neural retrieval models are still missing~\cite{voelske:2021}. Consequently, our axiomatic error analysis aims to answer the following research question:

\newcommand{\RQfive}{\begin{itemize}
    \item[\textbf{RQ5}] \emph{Can retrieval axioms explain why BM25 is better at effectiveness on Touch{\'e}~2020 than neural retrieval models?}
\end{itemize}}
\RQfive

\input{tables/lnc2-analysis}

\paragraph{Setup and Background}
We conduct our axiomatic analysis using the \texttt{ir\_axioms} framework~\cite{bondarenko:2022}.\footnote{\href{https://github.com/webis-de/ir\_axioms}{https://github.com/webis-de/ir\_axioms}} Because most axioms require theoretical preconditions that are rarely met in real-world datasets (e.g., requiring document pairs retrieved for the same query of identical length)~\cite{bondarenko:2022}, we first use synthetic document pairs derived from real documents and subsequently use real document pairs with the default length relaxation from \texttt{ir\_axioms}. Given more than 20~previously proposed retrieval axioms~\cite{bondarenko:2022}, we include a subset of all axioms related to document length, term frequency, and semantic similarity in our analysis. We focus on document length axioms following our observation that document length plays an important role in Touch{\'e}~2020, while we include term frequency and semantic similarity because they are the specialty of lexical and neural retrieval models. In all cases, we report the agreement in the percentage of the model under investigation with the preferences of an axiom as implemented in \texttt{ir\_axioms}.

\paragraph{Axiomatic Analysis on Synthetic Document Pairs}
Table~\ref{table-lnc2-analysis} shows the agreement of the tested models with the document length normalization axiom~LNC2 that (somewhat artificially) states that the relevance score of an $m$-times self-concatenation of a document should not be lower than the original document's relevance score~\cite{fang:2011}. We synthetically create document pairs that fulfill this precondition by randomly sampling 250~query--document pairs from the top-10 ranked results by all models under investigation. For each query--document pair, we create pairs for $m=1$, $2$, $3$, and $4$. We observe that BM25 almost perfectly agrees with the LNC2 axiom (agreement above~99\%), whereas neural models substantially violate LNC2, with TAS-B having the highest disagreement, which is an expected shortcoming of TAS-B as all documents are, independent of their length, represented by vectors of the same length.

\input{tables/axiomatic-analysis}

\paragraph{Axiomatic Analysis on Real Document Pairs}
Table~\ref{table-axiomatic-analysis} shows the results of our axiomatic analysis on all document pairs from the top-50 ranked results for each test query on both the original~(O) and the denoised~(+D) Touch{\'e}~2020 corpus. We report the term frequency axioms TFC1~\cite{fang:2004}, TFC3~\cite{fang:2011} (we leave out TFC2~\cite{fang:2011} because this axiom can only be applied on synthetic documents), and TDC~\cite{fang:2011}, the document length axioms LNC1~\cite{fang:2011} and TF-LNC~\cite{fang:2011}, and the semantic similarity axioms STMC1~\cite{fang:2006} and STMC2~\cite{fang:2006}. We observe that BM25 has the highest agreement with the term frequency axioms TFC1 and TFC3 which are more frequently violated by the other neural models. For the M-TDC, LNC1, and TF-LNC axioms, BM25 achieves only mediocre agreement. Similarly, BM25 does not agree well with the semantic similarity axioms STMC1 and STMC2, where neural models outperform BM25, for which this could be expected (BM25 alone suffers from vocabulary mismatch in contrast to neural models), which indicates that those axioms play a subordinate role on Touch{\'e}~2020.

\section{Discussion and Future Work}\label{sec:discussions}

Our systematic evaluation reveals the limitations of existing neural retrieval models for argument retrieval.
These limitations largely stem from (i)~the noise (short arguments) present within Touch{\'e}~2020 and (ii)~the nature of the task that ties relevance with argument quality.
Ensuring that neural models do not merely focus on the high-lexical overlap between the query and retrieved document remains a challenge.
To tackle this problem, it is critical to teach retrieval models potentially via further training, to identify documents that are not just lexically similar but semantically relevant. We leave it as future work to investigate strategies for updating the training loss function with regularization terms that penalize short documents in Touch{\'e}~2020, a concept borrowed from document length normalization \cite{singhal:1996}, to improve robustness in retrieval systems against noise present within document corpus.

Our evaluation also reveals that Touch{\'e}~2020 corpus is rather noisy (similar to real-world test collections) containing many low-quality arguments and a lot of unjudged documents. Noisy data can create several problems that lead to the drawing of false conclusions. As shown in this work, enhancing data quality through careful denoising and post-hoc judgments leads to substantial improvements in the effectiveness of all retrieval models. We hope the community adopts similar insights from our work and potentially evaluate future model effectiveness on our denoised and post-hoc relevance judged Touch{\'e}~2020 dataset is publicly available at \href{https://github.com/castorini/touche-error-analysis}{https://github.com/castorini/touche-error-analysis}.

\paragraph{Limitations} We acknowledge that our work is not perfect and contains limitations. In our work, we conduct an in-depth study of argument retrieval. TREC-COVID~\cite{roberts:2021}, a bio-medical dataset in the BEIR benchmark observes a similar spike in short document distribution, as a large number of documents in the corpus do not contain an abstract (i.e., body)~\cite{thakur:2021}. We leave it as future work, to similarly investigate denoising and black-box model evaluation on TREC-COVID. Similarly, in our work, we investigate only the retrieval model's effectiveness in the first-stage argument retrieval. We did not evaluate cross-encoders or neural models at the second, i.e., reranking stage, in argument retrieval. Lastly, in our work, we did not retrain any model due to the additional computation costs. In the future, we would like to explore training robust neural models and implementing document length normalization as a regularization objective to make neural models less sensitive against noisy short documents in Touch{\'e}~2020.

\section{Conclusion}
In this paper, we addressed the question of why neural models are subpar, compared to BM25, on the BEIR subset Touch{\'e} 2020, an argument retrieval task. 
To this end, we conducted a systematic error analysis and found that neural models often retrieve short and non-relevant arguments. 
To alleviate this issue, we enhanced data quality by filtering out noisy and short arguments in Touch{\'e}~2020 and included post-hoc judgments to fill up holes for a fair evaluation of all tested models.
Although our amendments improve the effectiveness of neural models by up to a margin of~0.52 in terms of nDCG@10 scores, they still lag behind BM25.
Coupled with our theoretical analysis, we highlight that all neural models violate the document length normalization LNC2 axiom, intuitively explainable as documents are mapped to equal-size vectors.
Addressing these shortcomings demands improved training strategies to adapt neural models for argument retrieval. 
Drawing insights from our findings, future work may focus on instructing models to favor longer and high-quality argumentative documents or to better support traditional retrieval axioms.

\begin{acks}
We thank our annotators for the post-hoc relevance judgments, Jack Lin for helping out reproducing DRAGON+, and Minghan Li for helping out reproducing CITADEL+. Our research was partially supported by the Natural Sciences and Engineering Research Council~(NSERC) of Canada; computational resources were provided by Compute Canada; by the DFG (German Research Foundation) through the project ``ACQuA~2.0: An\-swer\-ing Comparative Questions with Arguments'' (project~376430233) as part of the priority program ``RATIO: Robust Argumentation Machines'' (SPP~1999); by the European Union’s Horizon Europe research and innovation program under grant agreement No~101070014~(\href{https://openwebsearch.eu}{OpenWebSearch.EU}, \href{https://doi.org/10.3030/101070014}{https://doi.org/10.3030/101070014}); and by the Stiftung f{\"u}r Innovation in der Hochschullehre under the ``freiraum~2022'' call~(FRFMM-58/2022).
\end{acks}

\bibliographystyle{ACM-Reference-Format}
\balance
\bibliography{references}

\end{document}

%% file: tables/table-beir-effectiveness.tex
\begin{table*}[t]
    \caption{The motivation of our work: dense (left), multi-vector (top right), and sparse retrieval models (bottom right) are less effective than BM25 on the BEIR subset Touch{\'e}~2020; nDCG@10 scores taken from the referenced publications.}
    \setlength{\tabcolsep}{1pt}
    \renewcommand{\arraystretch}{1}
    {\begin{tabular}[t]{l@{}>{\centering\arraybackslash}p{10em}>{\centering\arraybackslash}p{4em}r}
        \toprule
        Model & \multicolumn{1}{c}{Reference} & Type & nDCG@10 \\
        \midrule
        BM25 (BEIR) & Thakur et al. \cite{thakur:2021} & lexical & \underline{\textbf{0.367}} \\ \midrule
        $\mathrm{E5_{large}}$ & Wang et al. \cite{wang:2022b} & dense & 0.272 \\
        BGE-large & Xiao et al. \cite{xiao:2023} & dense & 0.266 \\
        Promptagator & Dai et al. \cite{dai:2023} & dense & 0.266 \\
        DRAGON+  & Lin et al. \cite{lin:2023} & dense & 0.263 \\
        GTR-XXL & Ni et al. \cite{ni:2022} & dense & 0.256 \\
        GPL & Wang et al. \cite{wang:2022} & dense & 0.255 \\
        RocketQAv2 & Ren et al. \cite{ren:2021} & dense & 0.247 \\
        ANCE & Xiong et al. \cite{xiong:2021} & dense & 0.240 \\
        RetroMAE  & Xiao et al. \cite{xiao:2022} & dense & 0.237 \\
        Contriever & Izacard et al. \cite{izacard:2022} & dense & 0.204 \\
        TART-dual  & Asai et al. \cite{asai:2023} & dense & 0.201 \\
        TAS-B  & Hoffstätter et al. \cite{hofstatter:2021} & dense & 0.162 \\
        \bottomrule
    \end{tabular}}
    \hspace{10mm}
    \renewcommand{\arraystretch}{1.015}
    {\begin{tabular}[t]{l@{}>{\centering\arraybackslash}p{12em}>{\centering\arraybackslash}p{5em}r}
        \toprule
        Model & \multicolumn{1}{c}{Reference} & Type & nDCG@10 \\
        \midrule
        BM25 (BEIR)& Thakur et al. \cite{thakur:2021} & lexical & \underline{\textbf{0.367}} \\ \midrule
        CITADEL+ & Li et al. \cite{li:2023} & mult.-vec. & 0.342 \\
        XTR (XXL) & Lee et al. \cite{lee:2023} & mult.-vec. & 0.309 \\
        CITADEL & Li et al. \cite{li:2023} & mult.-vec. & 0.294 \\
        COIL-full & Gao et al. \cite{gao:2021} & mult.-vec. & 0.281 \\
        ColBERTv2 & Santharam et al. \cite{santhanam:2022} & mult.-vec. & 0.263 \\
        ColBERT & Khattab et al. \cite{khattab:2020} & mult.-vec. & 0.202 \\
        
        \midrule
        \addlinespace
        uniCOIL &  Lin et al. \cite{lin:2021} & sparse & 0.298 \\ 
        SPLADEv2 & Formal et al. \cite{formal:2021} & sparse &  0.272 \\
        SPLADE++ & Lassance et al. \cite{lassance:2022} & sparse & 0.244 \\
        DeepCT &  Dai et al. \cite{dai:2021} & sparse &  0.175 \\
        SPARTA & Zhao et al. \cite{zhao:2021} & sparse & 0.156 \\
        \addlinespace
        \bottomrule
    \end{tabular}}
    \label{table-beir-effectiveness}
\end{table*}

%% file: tables/retrieval-document-examples.tex
\begin{table}[t]
     \caption{Example of the top-ranked document for a randomly selected query showing that neural models may retrieve documents with a relevant conclusion~/~title (within the < >) but a non-relevant premise~/~body. \hlc[green!20]{Green}: relevant document; \hlc[red!15]{red}: non-relevant document.}
    \small
    {\begin{tabular}{p{8cm}}
    \multicolumn{1}{l}{Query (qid=5): \emph{Should social security be privatized?}} \\ \midrule
    \cellcolor{green!20}\textbf{BM25}: \cellcolor{green!20}$<$Social security should be privatized$>$ Social Security has serious issues [ ... ] First, privatization has a shaky track record. A 2004 report from the World Bank (http://wbln1018.worldbank.org) [ ... ] \\ \midrule
    \cellcolor{green!20}\textbf{CITADEL+}: \cellcolor{green!20}$<$Social security should be privatized$>$ - Social security is a complete joke. Although it was originally designed [ ... ] the young are forced to subsidize the old, a facet of socialism [ ... ] \\ \midrule
    \cellcolor{red!15}\textbf{SPLADEv2}: \cellcolor{red!15}$<$Social Security R.I.B Should be Privatized$>$ Thank you lannan13 for an invigorating debate. \\ \midrule
    \cellcolor{red!15}\textbf{DRAGON+}: \cellcolor{red!15}$<$Social Security R.I.B Should be Privatized$>$ Pass \\ \midrule
    \cellcolor{red!15}\textbf{Contriever}: \cellcolor{red!15}$<$Social Security R.I.B Should be Privatized$>$ Thank you lannan13 for an invigorating debate. \\ \midrule
    \cellcolor{red!15}\textbf{TAS-B}: \cellcolor{red!15}$<$Privatizing social security$>$ Social security is in crisis \\
    \bottomrule
    \end{tabular}}
   
    \label{tab:retrieval-documents-examples}
\end{table}

%% file: tables/error-rate-table.tex
\begin{table}[t]
    \caption{Error rates as the percentage of a model's top-$k$ results that are non-relevant (judgment of~0 or unjudged) and shorter than 20~words. Lower error rates are better.}
    \setlength{\tabcolsep}{2pt}
    \renewcommand{\arraystretch}{1}
    \small
    {\begin{tabular}{@{}l c c c c c c@{}}
    \toprule
    Model & \multicolumn{1}{c}{BM25} & \multicolumn{1}{c}{CITADEL+} & \multicolumn{1}{c}{SPLADEv2} & DRAGON+ & Contriever & TAS-B \\ 
    \cmidrule(l{-0.05em}r{0.5em}){1-1} \cmidrule(lr){2-2} \cmidrule(lr){3-3} \cmidrule(lr){4-4} \cmidrule(lr){5-7}
    \textbf{Top-1} & 0.0\% & 6.1\% & 22.4\% & 40.8\% & 55.1\% & 59.2\% \\ \midrule
    \textbf{Top-5} & 0.4\% & 3.3\% & 15.9\% & 32.7\% & 40.4\% & 59.2\% \\ \midrule
    \textbf{Top-10} & 0.8\% & 4.5\% & 14.6\% & 26.5\% & 35.9\% & 51.6\% \\ \bottomrule
\end{tabular}}\label{tab:error-rate-table}
\end{table}

%% file: tables/exp-summ-final.tex
\begin{table}[t]
    \caption{Example queries and Touch{\'e}~2020 documents: original and modified by replacing with a GPT-3.5 summary~\cite{ouyang:2022} or expanded by DocT5query queries~\cite{nogueira2019doc2query}. \hlc[green!20]{Green}: a relevant document; \hlc[red!15]{red}: a non-relevant document.}
    \small
    {\begin{tabular}{p{8cm}}
    \multicolumn{1}{l}{Query (qid=13): \emph{Can alternative energy effectively replace fossil fuels?}} \\ \midrule
    \textbf{\cellcolor{green!20}Original}: $\langle$fossil fuel$\rangle$ [ ... ] there are many alternatives to fossil fuel [ ... ] some of these alternatives are Nuclear fusion geothermal energy wind and solar power [ ... ] Nuclear fusion is a very effective way for one to create a mass amount of energy [...] \\ \midrule
    \cellcolor{green!20}\textbf{GPT-3.5 Summary}: \cellcolor{green!20}The argument presented is that there are many alternatives to fossil fuel and that these alternatives, such as nuclear fusion, geothermal energy, and solar and wind power, are both efficient and cost-effective. The argument emphasizes the need for a new and better source of energy [...]  \\ \midrule
    \multicolumn{1}{l}{Query (qid=2): \emph{Is vaping with e-cigarettes safe?}} \\ \midrule
    \textbf{Original}: \cellcolor{red!15}$\langle$Cigarettes should be banned$\rangle$ They are bad \\ \midrule
    \textbf{DocT5query}: \cellcolor{red!15}They are bad why are oohs swollen and puffy why do bad people make up names are narcotics bad why are morgans bad what is the reason they are bad are the oxen bad why are fish really bad for kids why do humans keep bad odours? are spiders bad [ ... ] \\ \bottomrule
    \end{tabular}}
    
     \label{tab:examples-summ-exp}%
\end{table}

%% file: tables/dataset-filteration.tex
\begin{table}[t]
 \caption{The Touch{\'e}~2020 dataset characteristics before and after denoising and post-hoc judgments. Reported are the total number of documents, the average document length, the number of queries, the number of relevance-judged documents, and the number of documents per relevance grade: non-relevant~(0), relevant~(1), and highly relevant~(2).}%
     \setlength{\tabcolsep}{6pt}
     \renewcommand{\arraystretch}{1}
     {\begin{tabular}{@{}l@{\quad}rrr@{}}
     \toprule
     \multicolumn{1}{c}{} & \textbf{Original} & \textbf{Denoised} & \textbf{Post-hoc} \\
     \cmidrule(l{0.1em}r{0.3em}){2-2}
     \cmidrule(l{0.1em}r{0.3em}){3-3}
     \cmidrule(l{0.2em}r{0.3em}){4-4}
 \textbf{\#~Documents} &  382,545 & 303,732 & 303,732 \\ \midrule
 \textbf{Avg.~length} & 293.5 & 358.7 & 358.7 \\ \midrule
 \textbf{\#~Queries} & 49 & 49 & 49 \\ \midrule
\textbf{\#~Judgments} &  2,214 & 1,785 & 2,849 \\ \midrule
\textbf{\#~Relevance~$=2$} & 636 & 620 (16$~\downarrow$) & 1,136 (516~$\uparrow$) \\
\textbf{\#~Relevance~$=1$} & 296 & 265 (31$~\downarrow$) & 576 (311~$\uparrow$) \\
\textbf{\#~Relevance~$=0$} & 1,282 & 900 (382$~\downarrow$) & 1,137 (237~$\uparrow$) \\ \bottomrule
    \end{tabular}}
     \label{tab:touche-stats}%
\end{table}

%% file: tables/filteration-results.tex
\begin{table}[t]
 \caption{Retrieval effectiveness as nDCG@10 and missing judgments as hole@10 on the original, denoised (cf.~Section~\ref{sec:denoising}), and post-hoc judged (cf.~Section~\ref{sec:post_hoc_judgements}) Touch{\'e}~2020 data showing that BM25 still outperforms the neural retrievers even after denoising and after post-hoc judgments.}
    \centering
    \setlength{\tabcolsep}{1pt} 
    \resizebox{0.47\textwidth}{!}
    {\begin{tabular}{@{}lcc@{\quad}cc@{\quad}cc@{}}
        \toprule
         &  \multicolumn{2}{c}{\textbf{Original}} & \multicolumn{2}{c}{\textbf{+ Denoised}} & \multicolumn{2}{c}{\textbf{++ Post-hoc}} \\
        \cmidrule(lr{1.4em}){2-3} \cmidrule(lr{1.4em}){4-5} \cmidrule(lr){6-7} 
        Model & {\small nDCG@10} & {\small hole@10} & \multicolumn{1}{c}{\small nDCG@10} & {\small hole@10} & \multicolumn{1}{c}{\small nDCG@10} & $\delta$ inc.\\ \midrule
        BM25 & \underline{\textbf{0.367}} & 61.6\% &  \underline{\textbf{0.467}} & 51.8\% & \underline{\textbf{0.785}} & $\triangle$~0.418 \\ \midrule
        CITADEL+ & 0.339 & 60.2\% &  0.362 & 62.5\% & 0.703 & $\triangle$~0.364 \\ \midrule
        SPLADEv2  & 0.272 & 66.3\% & 0.326 & 63.3\% & 0.679 & $\triangle$~0.407 \\ \midrule
        DRAGON+ & 0.249 & 69.2\% & 0.340 & 63.9\% & 0.718 & $\triangle$~0.469 \\
        Contriever & 0.205 & 71.4\% & 0.303 & 65.9\% & 0.650 & $\triangle$~0.445 \\ 
        TAS-B & 0.162 & 77.8\% & 0.306 & 67.5\% & 0.682 & $\triangle$~0.520 \\ \bottomrule
    \end{tabular}}   
    \label{tab:filtered-results}
\end{table}

%% file: tables/lnc2-analysis.tex
\begin{table}
\caption{Agreement (in \%) with the length normalization axiom~LNC2 when retrieving with (w/) or without (w/o) the title on Touch{\'e}~2020. BM25 agrees perfectly with LNC2.}
\setlength{\tabcolsep}{2.0pt}
\renewcommand{\arraystretch}{1.1}
\small
{\begin{tabular}{@{}lcccccc@{}}
    \toprule
     & BM25 & CITADEL+ & SPLADEv2 & DRAGON+ & Contriever & TAS-B \\
    \midrule
    w/\phantom{o} title & 99.6 & 75.3 & 60.6 & 39.2 & 41.8 & 35.2 \\
    w/o title & 99.5 & 79.1 & 68.2 & 39.5 & 40.8 & 38.9 \\ 
    \bottomrule
\end{tabular}
}
\label{table-lnc2-analysis}

\end{table}

%% file: tables/axiomatic-analysis.tex
\begin{table}
\caption{Agreement (in \%) with the term frequency, document length, and semantic similarity axioms for all tested models on the original~(O) and the denoised~(+D) Touch{\'e}~2020 data.}
\small
\setlength{\tabcolsep}{1.85pt}
\renewcommand{\arraystretch}{1}
{\begin{tabular}{@{}lccc@{\quad}cc@{\quad}cc@{}}
    \toprule
    \textbf{Model} & \multicolumn{3}{@{}c@{}}{\textbf{Term Frequency}} & \multicolumn{2}{@{}c@{\quad}}{\textbf{Doc. Length}} & \multicolumn{2}{@{}c@{}}{\textbf{Semantic Sim.}}\\

    \cmidrule(r{7pt}){2-4} \cmidrule(r{6pt}){5-6} \cmidrule{7-8}

    & TFC1 & TFC3 & M-TDC & LNC1 & TF-LNC  & STMC1 & STMC2\\
    \midrule
        
    BM25 (O) & 61.6 & 100.0 & 51.8 & 37.8 & 58.5 & 48.3 & 54.9 \\
    BM25 (+D) & 68.5 & 100.0 & 55.6 & 32.8 & 57.4 & 48.4 & 50.9 \\
    CITADEL+ (O)& 59.2 & 88.9 & 56.6 & 54.3 & 60.7 & 50.7 & 57.9 \\
    CITADEL+ (+D)& 62.6 & 72.7 & 47.6 & 56.1 & 57.1 & 51.0 & 57.8 \\
    Contriever (O) & 59.7 & 100.0 & 46.5 & 52.7 & 55.9 & 52.5 & 59.1 \\
    Contriever (+D) & 59.4 & 80.0 & 51.4 & 52.5 & 57.7 & 52.6 & 54.3 \\
    DRAGON+ (O) & 61.1 & 100.0 & 50.6 & 55.3 & 59.0 & 52.1 & 58.2 \\
    DRAGON+ (+D) & 63.2 & 92.3 & 54.7 & 53.1 & 55.4 & 52.2 & 54.5 \\
    SPLADEv2 (O) & 59.8 & 50.0 & 57.1 & 47.8 & 56.2 & 50.8 & 56.6 \\
    SPLADEv2 (+D) & 62.9 & 91.7 & 53.0 & 51.5 & 57.8 & 51.3 & 55.2 \\
    TAS-B (O) & 60.1 & 33.3 & 55.4 & 50.3 & 55.8 & 52.3 & 60.5 \\
    TAS-B (+D) & 62.2 & 33.3 & 50.6 & 54.0 & 53.1 & 52.4 & 54.2 \\

    \bottomrule
\end{tabular}}
\label{table-axiomatic-analysis}
\end{table}